\documentclass{cflowproc}

\usepackage{natbib}

\begin{document}

\shortauthors{KAISER \& BINNEY}     
\shorttitle{ENTROPY EVOLUTION IN CLUSTERS} 

\title{Entropy Evolution of the Gas in Cooling Flow Clusters}   

\author{Christian R. Kaiser,\affilmark{1}   
James J. Binney,\affilmark{2}                
}

\affil{1}{University of Southampton}   
\and                                
\affil{2}{University of Oxford}

\begin{abstract}
We emphasise the importance of the gas entropy in studying the
evolution of cluster gas evolving under the influence of radiative
cooling. On this basis, we develop an analytical model for this
evolution. We then show that the assumptions needed for such a model
are consistent with a numerical solution of the same equations. We
postulate that the passive cooling phase ends when the central gas
temperature falls to very low values. It follows a phase during which
an unspecified mechanism heats the cluster gas. We show that in such a
scenario the small number of clusters containing gas with temperatures
below about 1\,keV is simply a consequence of the radiative cooling.
\end{abstract}

\section{Introduction}

Various speakers at this meeting have shown overwhelming evidence that
to explain our observations of the hot gas in galaxy clusters we need
to invoke some form of non-gravitational heating of this gas. Various
mechanisms for heating the gas in galaxy clusters have been
proposed. All of these can be divide into two categories: Steady-state
or episodic heating scenarios. In the steady-state picture the heating
mechanism `knows' locally about the radiative cooling. Every bit of
energy radiated away by the gas at a given place inside the cluster is
replaced by an equal bit of energy at the same place by the heating
process. The cluster as whole never changes. For episodic heating this
tight spatial and temporal connection between cooling and heating is
broken. The cluster gas spends some time passively cooling and the
lost energy is then replenished during a heating period. The energy
input does not have to be distributed in this case. For example, the
heating of the cluster gas by an AGN outflow may initially only affect
the very centre of the cluster. However, such a localised release of
energy adds the further complication of having to subsequently
distribute the energy throughout at least the inner volume of the
cluster. This distribution of energy cannot be achieved by a fully
convective flow as this would imply a negative entropy gradient in the
cluster which has never been observed.

In this contribution we are not investigating any heating
mechanisms. We simply assume that the cluster gas is heated in an
episodic fashion, possibly by AGN outflows. In between heating phases
the cluster gas cools passively until the temperature of the gas at
the cluster centre vanishes. At this point the heating mechanism is
triggered once more. In the following we solve the hydrodynamical
equations governing the evolution of the cluster gas during the
passive cooling periods. We will show that the properties of the
cluster gas predicted by our model are consistent with
observations. In particular we demonstrate that our model can explain
why we find very few clusters containing gas at temperatures below
about 1\,keV. 

In section \ref{entropy} we motivate our approach of concentrating on
the entropy of the cluster gas rather than its internal
energy. Section \ref{analytics} outlines the development of a simple
analytical model describing the evolution of the cluster gas. The
assumptions for the analytical model are tested against a numerical
solution of the same equations in section \ref{numerics}. Finally, in
section \ref{cold} we use the numerical approach to demonstrate that
passive radiative cooling alone can explain the apparent paucity of
detections of clusters containing very cold gas.

\section{Why is entropy important?}
\label{entropy}

In the following we give two reasons why the entropy of a gas is
an important quantity to study in galaxy clusters. The first is a
significant simplification of the hydrodynamic equations
describing the cooling flow gas. Secondly, observations of cluster
gas reveal a simple power-law relation between the gas' entropy and
its mass.

\subsection{Comparison with energy}

The properties of the hot gas in galaxy clusters evolving under the
influence of radiative cooling are traditionally described by a set of
three hydrodynamical equations: Conservation of mass, momentum and
energy \citep[e.g.][]{cs88}. Usually these equations are then solved
numerically and the resulting inflow of gas towards the cluster centre
due to radiative cooling is referred to as a cooling flow. It is
intuitive to use the conservation of energy equation as the rate of
change of the internal energy of the gas due to radiative cooling is
very simple, i.e.
\begin{equation}
\frac{\partial E_{\rm rad}}{\partial t} \propto \rho ^2 \Lambda
(T),
\end{equation}
where $\rho$ is the mass density and $\Lambda (T)$ is the
temperature-dependent cooling function. However, the total rate of
change of the internal energy is much more complicated. As the gas
cools, it is compressed by the somewhat hotter gas further out in
the cluster. This adiabatic work increases the internal energy.
Radiative cooling is a strong function of the gas density. Thus
cooling is most important in the cluster centre and the continued
compression of the gas there leads to a gas flow inwards. On the
way in the gas looses gravitational potential energy. Other ways
in which the gas may gain or loose energy include the conversion
of kinetic energy of the gas flow and thermal conduction. All
these processes make the equation of energy difficult to solve.

In fact, we can formulate an equivalent, but considerably simpler
set of equations to describe the evolution of the cluster gas. For
this, we consider a variant of the entropy of the gas defined as
\begin{equation}
S = N k_{\rm B} \left[ {\rm const.} +\frac{3}{2} \ln \left( p \rho
^{-5/3} \right) \right],
\end{equation}
where $N$ is the number of gas atoms, $k_{\rm B}$ is the Boltzmann
constant and $p$ is the gas pressure. Here we have assumed a
monatomic gas with adiabatic index equal to $5/3$. Entropy itself
is somewhat cumbersome to work with and so we define the `entropy
index' as
\begin{equation}
\sigma = p \rho^{-5/3}.
\end{equation}

Entropy and therefore the entropy index does not change for
adiabatic compression of the gas. If we further assume that all
gas motions are subsonic, i.e. the kinetic energy of the gas is
small compared to other forms of energy, and that thermal
conduction is suppressed, then the only process changing the gas
entropy is radiation. It is straightforward to show that
\begin{equation}
\dot{\sigma} = - \frac{2}{3} \rho^{1/3} \Lambda
(T).
\label{dsigma}
\end{equation}
We will see in the following that using this simple expression instead
of the conservation of energy equation leads to an analytical solution
of the hydrostatic equations.

\subsection{Entropy index -- mass relationship}
\label{entmass}

\begin{figure}
\plotone{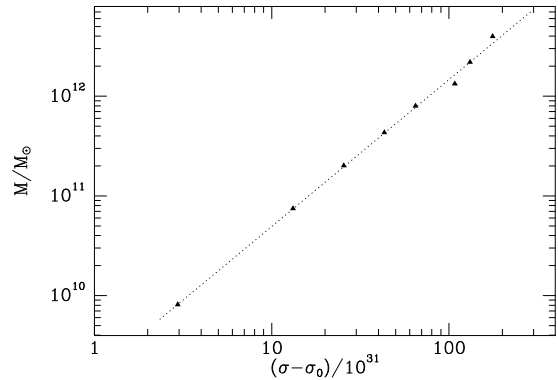} 
\figcaption{Distribution of gas mass as function
of entropy index, $\sigma$, for the Hydra cluster. Triangles show
observational results from \citet{dnm01}. The line shows the
best-fitting relation of the form given by equation (\ref{power}). The
units of $\sigma$ are cm$^4$\,g$^{-2/3}$\,s$^{-2}$.
\label{hydra}}
\end{figure}

\begin{figure}
\plotone{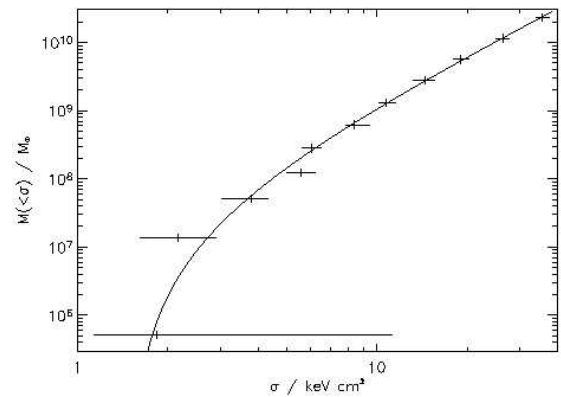} 
\figcaption{Distribution of gas mass as function
of entropy index, $\sigma$, for the Virgo cluster. Data points show
observational results from \citet{mbf02}. The line shows the
best-fitting relation of the form given by equation (\ref{power}).
\label{virgo}}
\end{figure}

\citet{kb02} pointed out that in the Hydra cluster the cumulative mass
of gas with an entropy index less than a given $\sigma$ is a simple
power-law function of $\sigma$ itself, i.e.
\begin{equation}
M(< \sigma ) = A \left( \sigma - \sigma _0 \right) ^{\epsilon},
\label{power}
\end{equation}
where $A$ is a constant and $\sigma _0$ is the entropy index of the
gas with the lowest entropy in the entire atmosphere located at the
centre of the cluster. Figure \ref{hydra} shows how closely a
power-law of this form fits the observations. Another example is the
Virgo cluster \citep[see Figure \ref{virgo}; ][]{ck03a}. More clusters
for which a similar relation may very well provide an excellent fit to
the data can be found in the contributions to these proceedings by
Horner et al. and Donahue.

A parcel of gas with a given entropy index at a specific time $t$ has
a well-defined mass. This mass will not change under the influence of
radiative cooling. Therefore at time $t+{\rm d}t$ the parcel of gas
will have a lower entropy index, but the mass of the gas in the parcel
with this new value of $\sigma$ will not have changed. Thus we can
suspect that the power-law relation between $M(<\sigma)$ and $\sigma -
\sigma _0$ may persist under the influence of radiative cooling. This
was in fact confirmed by \citet[][see also section
\ref{numerics}]{kb02} by numerical integration of the hydrodynamical
equations. An analytical argument for this behaviour will be presented
in a forthcoming paper.

\section{Analytical model for evolving cooling flows}
\label{analytics}

We use equation (\ref{dsigma}) giving the rate of change of the
entropy index to replace the conservation of energy equation. The
conservation of momentum equation remains unchanged, i.e.
\begin{equation}
\frac{\partial p}{\partial r} = - \rho \frac{\partial
\Phi}{\partial r} = - \left( \frac{p}{\sigma} \right)^{3/5}
\frac{\partial \Phi}{\partial r},
\label{hydrostatic}
\end{equation}
where $\Phi$ is the gravitational potential. As usual we assume
that $\Phi$ is generated by the dark matter halo of the cluster
and remains unchanged by the motion of the gas. 

We now assume a cooling function $\Lambda (T)$ appropriate for
radiative cooling due to pure bremsstrahlung. This implies $\Lambda
(T) \propto \sqrt{T}$. Clearly this is a strong simplification as it
neglects any cooling due to line emission. Even for a gas of vanishing
metallicity line emission will change the cooling function
significantly for gas with temperatures below about 2\,keV. However,
above 2\,keV this approximation of the cooling function is
reasonable. Due to the very efficient cooling at low temperatures most
of the cluster gas will spend only very little time with a temperature
below this threshold. With this we find
\begin{equation}
\dot{\sigma} \propto p^{2/5} \sigma ^{1/10}.
\end{equation}
Clearly the rate of entropy loss is only a weak function of the
changing gas properties. Therefore we set $p^{2/5} \sigma ^{1/10}
\approx$ const. Apart from the factor $\sigma ^{1/10}$, this is
equivalent to the assumption of isobaric cooling roughly consistent
with the fact that the cooling timescale is long compared to the
dynamical timescale of the gas.

Our final assumption in order to facilitate an analytical solution is
that of a uniform entropy index, $\sigma _{\rm i}$, of all the gas
throughout the cluster atmosphere at $t=0$. Again this may well be a
gross oversimplification of the real situation as the cluster gas is
presumably accumulated from a number of dark matter subhaloes merging
to form the final cluster halo. There is no reason why the gas
contained in these subhaloes should have the same entropy. However,
any heating process affecting the entire cluster will eventually lead
to a uniform entropy throughout the cluster atmosphere. Whether such a
comprehensive heating phase has taken place or not is of course not
clear. However, the assumption is crucial to obtaining an analytical
solution.

With these assumptions we can solve equations (\ref{dsigma}) and
(\ref{hydrostatic}) to find
\begin{equation}
\sigma = \left( \sigma _{\rm i}^{8/5} - c_0 \Phi t \right) ^{5/8}
\end{equation}
and 
\begin{equation}
p = \left[ \left( \sigma _{\rm i} - \sigma \right) \frac{c_1}{t}
\right] ^{5/2}.
\end{equation}
Here, $c_0$ and $c_1$ are constants and the other thermodynamic
properties of th gas may be derived from the above two expressions by
assuming ideal gas conditions. Also note that the solution does not
depend on the form of the gravitational potential $\Phi$.

\begin{figure}
\plotone{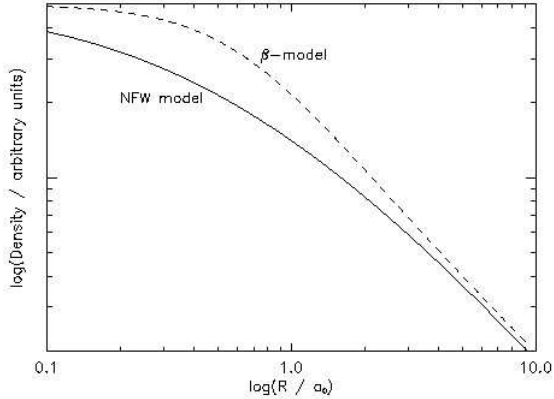} 
\figcaption{Density distribution of the cluster
gas in the analytical model for two different gravitational
potentials.
\label{density}}
\end{figure}

\begin{figure}
\plotone{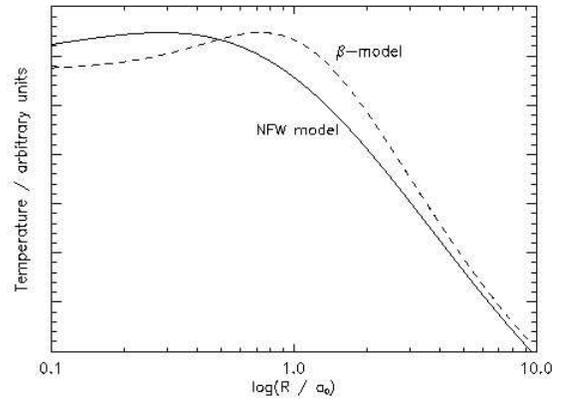} 
\figcaption{Temperature distribution of the
cluster gas in the analytical model for two different gravitational
potentials.
\label{temperature}}
\end{figure}

Figures \ref{density} and \ref{temperature} illustrate the density and
temperature distributions in the cluster atmosphere predicted by the
analytical model. As expected, assuming a $\beta$-profile for the
gravitational potential results in the formation of a uniform density
core while a NFW-profile \citep{nfw96} results in a steeper central
density distribution. The temperature distributions show an off-centre
peak for both distributions, but the $\beta$-profile also results in a
constant temperature core. This general shape with a constant
temperature followed by a peak further out is reminiscent of the
generic temperature profile found empirically by \citet{asf01}
studying a sample of cooling flow clusters.

\begin{figure}
\plotone{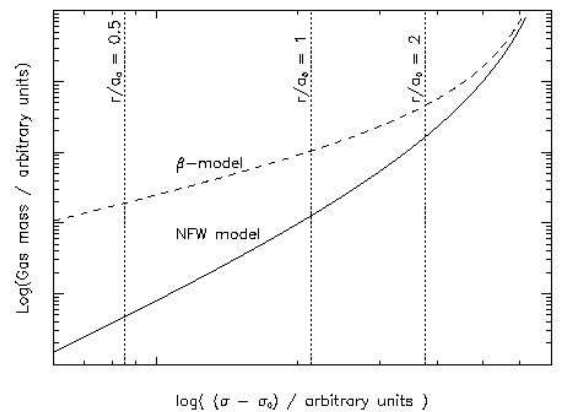} 
\figcaption{Cumulative mass, $M(<\sigma)$, as a
function of entropy index as predicted by the analytical model. The
vertical lines show the location of the gas within the cluster for the
model with a gravitational potential described by the $\beta$-profile.
\label{sigma_m}}
\end{figure}

Finally, Figure \ref{sigma_m} illustrates that, over a large range in
entropy index and radius, the analytical model predicts a power-law
relation between the gas mass and the entropy index as described by
equation (\ref{power}). The slope of the power-law, $\epsilon$, in the
inner part of the relation is roughly equal to 3 for the NFW-profile
and $3/2$ for the $\beta$-profile independent of the value of
$\beta$. Given that the relation steepens for increasing radius, the
latter value is closer to the slopes (Hydra: $\epsilon \sim 1.5$,
Virgo: $\epsilon \sim 2.3$) found empirically from the data shown in
Figures \ref{hydra} and \ref{virgo}.

\section{Numerical approach}
\label{numerics}

The results from the analytical model presented above may arise as a
direct consequence of the great simplifications we employed in
deriving this solution. In order to test this possibility, we have
also solve equations (\ref{dsigma}) and (\ref{hydrostatic}) by
numerical integration. Without the assumptions we made above, we need
a third equation to close the system. Here we use a variant of the
equation of mass conservation in the form
\begin{equation}
\frac{\partial M}{\partial r} = \frac{\partial M}{\partial \sigma}
\frac{\partial \sigma}{\partial r} = 4 \pi r^2 \rho.
\label{mass}
\end{equation}
We start the integration with the gas distribution derived from the
X-ray observations of the Hydra cluster given in \citet{dnm01}. These
data provide us with the initial values for the derivative $\partial M
/ \partial \sigma$. We then use equation (\ref{dsigma}) to evolve the
entropy index through a small timestep ${\rm d} t$. We calculate the
new derivative using the fact that the mass of gas with a given
entropy index does not change (see section \ref{entmass}). Now it is
possible to numerically solve equations (\ref{dsigma}) and
(\ref{mass}). We then repeat the whole procedure for the next
timestep. The numerical solution also includes a more realistic
cooling function incorporating line emission. Details of the
calculations are given in \citet{kb02}.

\begin{figure}
\plotone{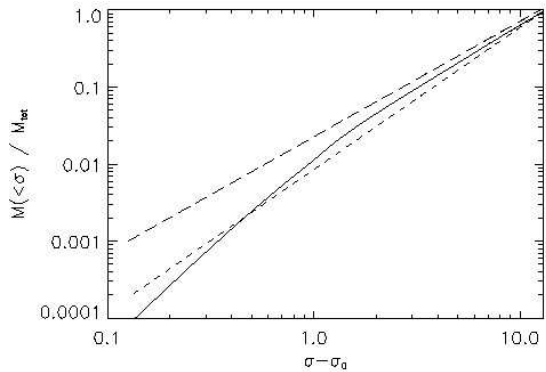} 
\figcaption{Entropy index -- mass relation
from the numerical solution. The long-dashed line shows the relation
at the beginning of the integration approximating the observational
data obtained in the Hydra cluster (see Figure \ref{hydra}). The
relation at the end of the calculation when the central entropy index
approaches zero is shown by the solid line. The short-dashed line
shows the power-law relation of the form of equation (\ref{power})
that provides the best approximation to the solid line.
\label{numeric_msig}}
\end{figure}

We continue the calculation until the entropy index and the
temperature of the gas at the cluster centre are equal to zero. Figure
\ref{numeric_msig} demonstrates that the entropy index -- mass
relationship continues to be well approximated by a power-law
throughout the calculation. The inclusion of radiative cooling due to
line emission does not change this relationship.

\begin{figure}
\plotone{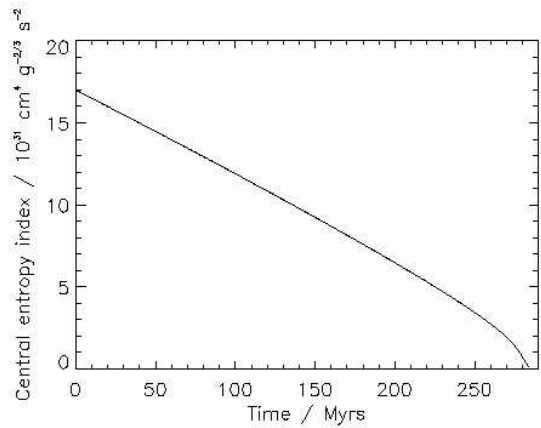} 
\figcaption{The evolution of the entropy
index of the gas at the cluster centre in the numerical solution.
\label{central_entropy}}
\end{figure}

The entropy evolution is strongest at the cluster centre as the
pressure is highest there. Therefore we expect the deviations of the
entropy evolution from a strictly linear behaviour to be strongest in
the cluster centre. Figure \ref{central_entropy} shows the evolution
of the entropy index of the gas at the cluster centre. Except for very
late times the graph shows that this evolution is linear with
time. Thus we conclude that our assumption for the analytical model of
a linear evolution of the entropy index throughout the cluster is a
good approximation to the real situation. 

\section{Why we do not detect cold gas in clusters}
\label{cold}

\begin{figure}
\plotone{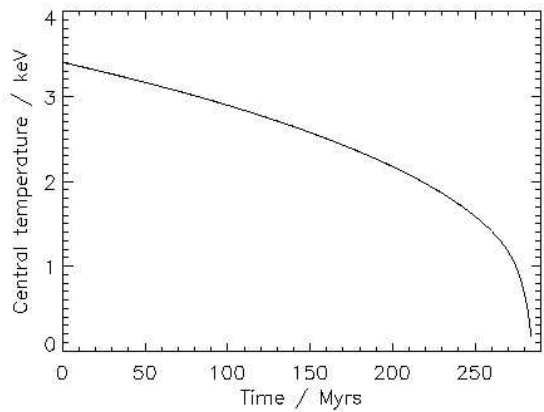} 
\figcaption{The evolution of the
temperature of the gas at the cluster centre in the numerical
solution.
\label{central_temperature}}
\end{figure}

A further prediction of the model in both its analytical or numerical
form is the strongly accelerating evolution of the gas temperature in
the cluster centre. Figure \ref{central_temperature} shows the
evolution of the gas temperature at the cluster centre. By
construction this is the coldest gas in the entire cluster. The
temperature decrease only gradually for a few $10^8$\,years to about
2\,keV. Only then the temperature drops rapidly to below 1\,keV. This
rapid drop is enhanced by the increasing importance of cooling due to
line emission at about this temperature, but the drop would also occur
in the absence of line cooling. The important point is that the lowest
gas temperature in the cluster exists only for a very short fraction
of the cluster evolution timescale. Furthermore, the volume of gas
with temperatures below 1\,keV or even 2\,keV is small. In other
words, the total emission measure of gas with temperatures below a
given low threshold is only a small fraction of the total emission
measure of the cluster.

\begin{figure}
\plotone{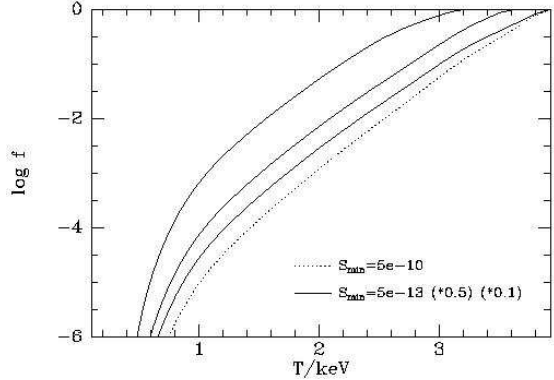} 
\figcaption{The fraction $f$ of clusters in a
flux-limited sample that contain a detectable amount of gas with
temperatures below a threshold $T$ as a function of $T$. The lines
show the fraction $f$ for various flux limits used to detect the cold
gas in follow up observations to the original survey. Dotted line: The
same flux limit as the original survey. Solid lines from bottom to
top: 1000, 2000 and $10^4$ times lower flux limits compared to the
original survey.
\label{fraction}}
\end{figure}

Consider a sample of clusters with a given lower flux limit. Assume
that every cluster in the sample has the same properties as the Hydra
cluster at $t=0$, but is observed at a random time in its
evolution. We can then use the numerical model to estimate the
fraction $f$ of clusters in this sample which contain a detectable
amount of gas at temperatures below a given threshold. Figure
\ref{fraction} presents the results of such an estimation. Even if the
original survey is followed up by much more sensitive observations to
search for cold gas, the fraction of clusters with detectable amounts
of gas with temperatures below 1\,keV is very small. It is therefore
not surprising that only very few clusters containing cold gas have
been found.

\section{Summary}

In this contribution we studied the evolution of the hot gas in galaxy
clusters under the influence of radiative cooling. By concentrating on
the entropy of the gas, we were able to find an analytical solution
for the relevant hydrodynamical equations. This solution assumes that
\begin{itemize}
\item the radiative cooling is due to pure bremsstrahlung.
\item the evolution of the entropy index of a given parcel of gas is
linear in time.
\item the entropy index has a uniform value throughout the cluster
atmosphere at $t=0$.
\end{itemize}
With the help of a numerical integration of the same basic equations
we showed that all of these assumptions are reasonable.

Both the analytical model and the numerical solution predict a simple
power-law relationship between the entropy index and the gas mass
consistent with observations. The accelerating temperature evolution
of the model can explain why we do not observe many clusters with very
cold gas at their centres. However, this result depends crucially on
the triggering of a heating mechanism for the cluster gas at the time
when the central temperature in the cluster approaches very low
values. Without such a heating mechanism, this model encounters the
usual problem that cold gas would rapidly accumulate at the cluster
centre. How the required heating mechanism works is not
clear. However, the arrival of cold gas at the cluster centre at the
time when the mechanism must be triggered may point to a close
connection between these two events.

\acknowledgements
CRK would like to thank PPARC for financial support.

\bibliography{crk}
\bibliographystyle{apj}

\end{document}